\documentclass[sigconf]{acmart}
\usepackage{hhline}
\usepackage{xcolor}
\usepackage{colortbl}
\usepackage{threeparttable}
\usepackage{subfig}
\usepackage{comment}
\usepackage{enumitem}
\usepackage[ruled,vlined]{algorithm2e}
\usepackage{tikz}
\newcommand*\circled[1]{\tikz[baseline=(char.base)]{
            \node[shape=circle,fill,inner sep=1.3pt] (char) {\textcolor{white}{#1}};}}

\newcommand{\fn}[1]{{\tt\small #1}}
\AtBeginDocument{%
  \providecommand\BibTeX{{%
    \normalfont B\kern-0.5em{\scshape i\kern-0.25em b}\kern-0.8em\TeX}}}

\copyrightyear{2023}  
\acmYear{2023} 
\setcopyright{rightsretained} 
\acmConference[HPDC '23]{Proceedings of the 32nd International Symposium on High-Performance Parallel and Distributed Computing}{June 16--23, 2023}{Orlando, FL, USA}
\acmBooktitle{Proceedings of the 32nd International Symposium on High-Performance Parallel and Distributed Computing (HPDC '23), June 16--23, 2023, Orlando, FL, USA}
\acmDOI{10.1145/3588195.3595955}
\acmISBN{979-8-4007-0155-9/23/06}

\renewcommand\footnotetextcopyrightpermission[1]{}
\begin{document}

\title{Accelerating MPI Collectives with Process-in-Process-based Multi-object Techniques}

\author{Jiajun Huang}
\email{jhuan380@ucr.edu}
\orcid{0000-0001-5092-3987}
\affiliation{%
  \institution{University of California, Riverside}
  \city{Riverside}
  \country{USA}
}

\author{Kaiming Ouyang}
\email{kouyang@nvidia.com}
\orcid{0000-0002-4775-1835}
\affiliation{%
  \institution{NVIDIA Corporation}
  \city{Santa Clara}
  \country{USA}
}

\author{Yujia Zhai}
\email{yzhai015@ucr.edu}
\orcid{0000-0002-2688-8058}
\affiliation{%
  \institution{University of California, Riverside}
  \city{Riverside}
  \country{USA}
}

\author{Jinyang Liu}
\email{jliu447@ucr.edu}
\orcid{0000-0003-0177-502X}
\affiliation{%
  \institution{University of California, Riverside}
  \city{Riverside}
  \country{USA}
}

\author{Min Si}
\email{msi@fb.com}
\orcid{0000-0002-0208-096X}
\affiliation{%
  \institution{Meta Platforms, Inc.}
  \city{Menlo Park}
  \country{USA}
}

\author{Ken Raffenetti}
\email{raffenet@anl.gov}
\orcid{0009-0003-4705-2713}
\affiliation{%
  \institution{Argonne National Laboratory}
  \city{Lemont}
  \country{USA}
}

\author{Hui Zhou}
\email{zhouh@anl.gov}
\orcid{0000-0002-4422-2911}
\affiliation{%
  \institution{Argonne National Laboratory}
  \city{Lemont}
  \country{USA}
}

\author{Atsushi Hori}
\email{ahori@nii.ac.jp}
\orcid{0000-0002-7010-8098}
\affiliation{%
  \institution{National Institute of Informatics}
  \city{Tokyo}
  \country{Japan}
}

\author{Zizhong Chen}
\email{chen@cs.ucr.edu}
\orcid{0000-0003-2578-4940}
\affiliation{%
  \institution{University of California, Riverside}
  \city{Riverside}
  \country{USA}
}

\author{Yanfei Guo}
\email{yguo@anl.gov}
\orcid{0000-0002-3731-5423}
\affiliation{%
  \institution{Argonne National Laboratory}
  \city{Lemont}
  \country{USA}
}

\author{Rajeev Thakur}
\email{thakur@anl.gov}
\orcid{0000-0002-5532-3048}
\affiliation{%
  \institution{Argonne National Laboratory}
  \city{Lemont}
  \country{USA}
}

\begin{abstract}

In the exascale computing era, optimizing MPI collective performance in high-performance computing (HPC) applications is critical. Current algorithms face performance degradation due to system call overhead, page faults, or data-copy latency, affecting HPC applications' efficiency and scalability. To address these issues, we propose PiP-MColl, a Process-in-Process-based Multi-object Inter-process MPI Collective design that maximizes small message MPI collective performance at scale. PiP-MColl features efficient multiple sender and receiver collective algorithms and leverages Process-in-Process shared memory techniques to eliminate unnecessary system call, page fault overhead, and extra data copy, improving intra- and inter-node message rate and throughput. Our design also boosts performance for larger messages, resulting in comprehensive improvement for various message sizes. Experimental results show that PiP-MColl outperforms popular MPI libraries, including OpenMPI, MVAPICH2, and Intel MPI, by up to 4.6X for MPI collectives like MPI\_Scatter and MPI\_Allgather.

\end{abstract}


\maketitle

\vspace{-2mm}

\section{Introduction}

Researchers have utilized data compression techniques to accelerate MPI collectives for large messages \cite{huang2023ccoll}. Kernel-assisted data copy approaches have been long demonstrated effective to speed-up intra-node communications. 
In \cite{chakraborty2017contention}, Chakraborty proposed Cross Memory Attach (CMA) which allows efficient intra-node communication without introducing redundant data copy. These approaches, however, suffer from significant performance degradation for small or medium-message collectives due to the overhead of expensive system calls and page faults. Parsons et al. demonstrated efficient MPI collective algorithms with a POSIX shared-memory (POSIX-SHMEM) \cite{parsons2014accelerating} multisender design. However, POSIX-SHMEM can limit the efficiency of algorithms that are unable to achieve high performance for medium- and large-message collective communication due to the inherent double copy overhead.

In addition to these data copy methods, researchers have also explored techniques that leverage data sharing based shared-memory (SHMEM) to reduce intra-node data transfer overhead. Hashmi et al. proposed strategies to reduce shared address space and accelerate \fn{MPI\_Allreduce} and \fn{MPI\_Reduce} communication using the interprocess SHMEM capability of XPMEM \cite{hashmi2018designing}. Nevertheless, XPMEM has system call overhead for buffer expose and attachment which limits its performance in small- and medium-messages. Apart from these previous shared-memory approaches, Hori et al. proposed Process-in-Process (PiP), a programming environment that allows all processes on a node to be loaded into the same virtual memory space and enables them to access each other's private memory as if they were threads in userspace \cite{hori2018process}. PiP is able to facilitate aforementioned extra copy and expensive system-related overhead.  A direct application of PiP in current MPI collectives, however, is unable to fully saturate the network bandwidth and is limited by the potential negative impact of unnecessary process synchronization on overall performance. To address these limitations, we introduce PiP-MColl, a Process-in-Process-based multi-object interprocess MPI collective design that maximizes intra- and inter-node message rate and network throughput.

\vspace{-0.55cm}

\section{Designs and optimizations}

An explicit multi-objective design achieves higher message rate and network throughput compared to previous single-object techniques for internode communication on small messages. This motivates the performance-oriented designs of our PiP-MColl collective algorithms. We present our algorithmic designs using \fn{MPI\_Allgather}. Traditionally, for small messages, the Bruck algorithm is used for non-power-of-two cases and the recursive doubling algorithm for power-of-two cases. To achieve high-performance in our allgather routine, we first design PiP-MColl allgather algorithm for small message sizes, which can be described as follows: \circled{1} We begin by performing intranode gather to the local root process. Local processes perform \fn{MPI\_Gather} to gather data into the local root process destination buffer $A_{d}$. \circled{2} Next, we initialize parameters. The multi-object Bruck algorithm step is initialized as $S_{p} = 1$ and the base of the multi-object Bruck algorithm as $B_{k} = P + 1$. \circled{3} We find the paired source and destination process. Each process sets $N_{offset} = (R_{l} + 1) * S_{p}$ and finds the paired source node $N_{src} = (N_{id} + N_{offset}) \% N$ and destination node $N_{dst} = (N_{id} - N_{offset}) \% N$. The paired source process rank is $N_{src} * N + R_{l}$, and the destination process rank is $N_{dst} * N + R_{l}$. \circled{4} We then perform send and receive. We define $C_{b}$ as the number of bytes received from each process in allgather and $A_{d}$ as the starting address of the destination buffer of the local root process. Each process sends $C_{b} * S_{p}$ bytes from the local root process buffer to the destination process and receives $C_{b} * S_{p}$ bytes from the source process into address $A_{d} + C_{b} * S_{p} * (R_{l} + 1)$. For each process, we update $S_{p} = S_{p} * B_{k}$. If $S_{p}$ is less than or equal to $\frac{N}{B_{k}}$, we repeat steps \circled{3} to \circled{4}. If not, we proceed to step \circled{5}. \circled{5} If $N$ is not a power of $B_{k}$, we have remaining $N - S_{p}$ nodes for the final step. Each process takes $Rem = Max(Min(S_{p}, N - S_{p} * R_{l}), 0)$ remainder. If $Rem > 0$, the process will send and receive $Rem * C_{b}$ bytes from the paired destination and source process. \circled{6} Finally, the local root process shifts data into the correct sequence and broadcasts to other processes.

\vspace{-2mm}

\section{Experimental Results}\label{exp-setup-sec}

\begin{figure}[ht]
	\centering
    \vspace{-4mm}
    \scalebox{0.7}{
	\includegraphics[width=\linewidth]{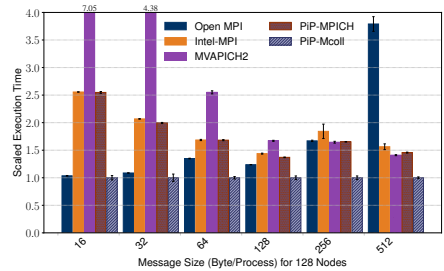}
    }
    \vspace{-7mm}
	\caption{MPI\_Scatter performance with small messages.}
	\label{fig-scatter-128-small}
    \vspace{-3mm}
\end{figure}

In our experiments, we use a 128-node cluster with 18 processes per node, resulting in a total of 2304 processes. Each node is equipped with two Intel Xeon E5-2695v4 Broadwell processors, yielding a total of 36 cores per node. The nodes are connected through the Intel OPA interconnect, with a maximum message rate of 97 million per second and a bandwidth of 100 Gbps. We evaluate the performance of the \fn{MPI\_Scatter} function for small message sizes. Figure \ref{fig-scatter-128-small} illustrates the scatter performance for small message sizes per process (i.e., overall data size on the root process is $M_{size} * \#process$) on a 128-node cluster with 18 processes on each node. To better distinguish performance differences between the MPI implementations, we have excluded execution times that are more than 4 times larger than that of PiP-Mcoll. As shown in the figure, PiP-MColl consistently outperforms the other MPI implementations, achieving the best speedup of 65\% when the message size is 256 bytes. This demonstrates the effectiveness of our multiobject design for small message sizes, as it maximizes the message rate and results in higher performance than other MPI implementations.

\begin{figure}[!htb]
	\centering
 \scalebox{0.7}{
	\includegraphics[width=\linewidth]{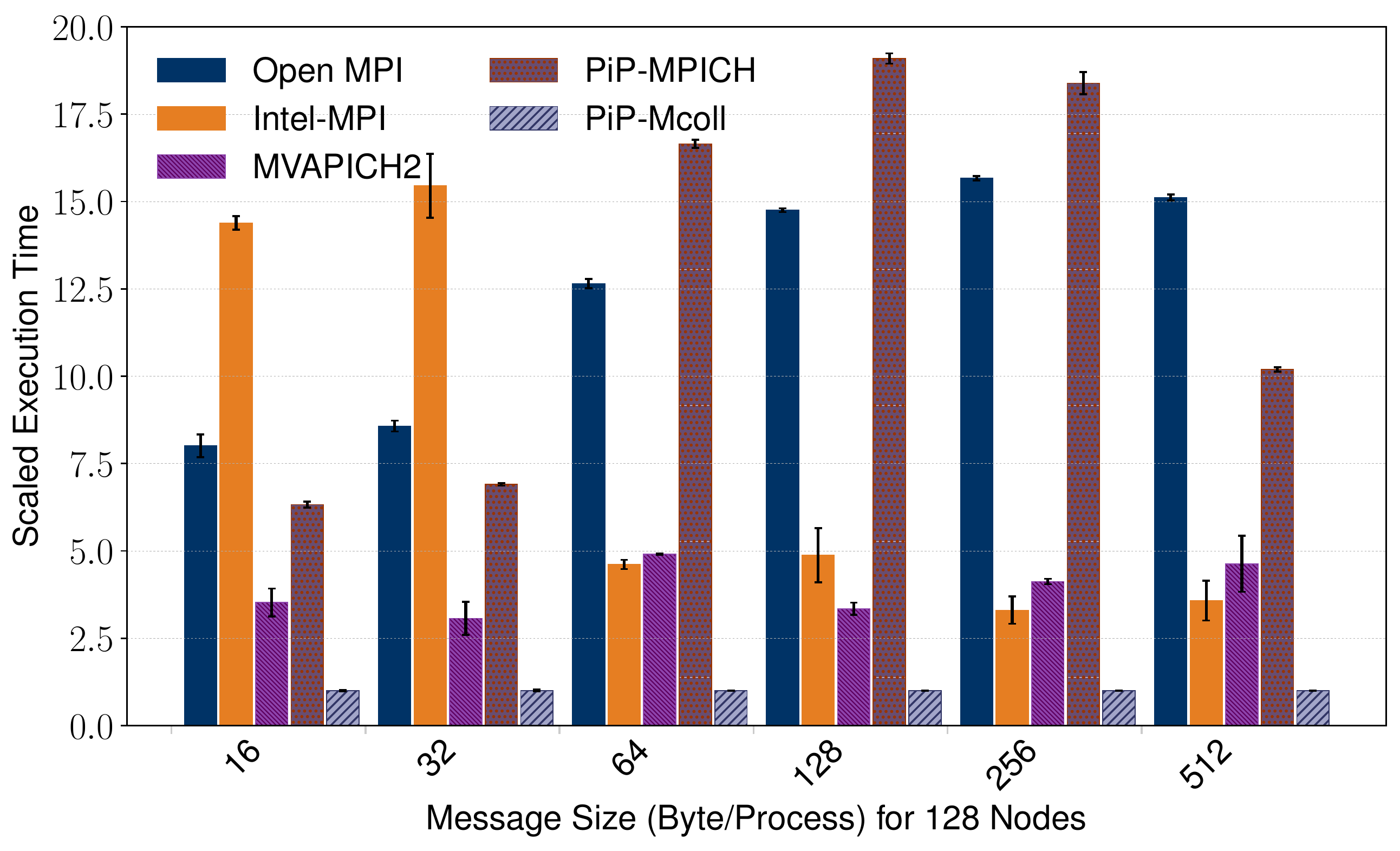}
  }
    \vspace{-3mm}
	\caption{MPI\_Allgather performance with small messages.}
	\label{fig-allgather-128-small}
    \vspace{-1mm}
\end{figure}

On the other hand, Figure \ref{fig-allgather-128-small} shows the \fn{MPI\_Allgather} performance with small message sizes from 16 B to 512 B on 128 Xeon Broadwell nodes. Theoretically, \fn{MPI\_Allgather} generates more data movements compared to other two MPI collectives, as processes receive more data than they send, which benefits the most from PiP-MColl. From the experimental data, we find that PiP-MColl outperforms other MPI implementations in all cases. Similarly, our multi-object design brings a noticeable performance improvement for small messages (i.e., 64 B), where PiP-Mcoll is over 4.6X as fast as the fastest MPI implementation. We also observe that our baseline (PiP-MPICH) sometimes has the worst performance among all the MPI implementations. This is due to the synchronization overhead inside PiP, which requires message size synchronization before communications. 


To conclude, PiP-MColl, our proposed Process-in-Process-based MPI Collective design, effectively optimizes MPI collective performance in HPC applications by addressing system call overhead, page faults, and data-copy latency issues. The result is a comprehensive improvement in message rate and throughput for various sizes, outperforming popular MPI libraries by up to 4.6X.
\vspace{-2mm}
\section*{Acknowledgment}
This research was supported by the Exascale Computing Project (17-SC-20-SC), a
 collaborative effort of the U.S.\ Department of Energy Office of Science and
the National Nuclear Security Administration, under contract DE-AC02-06CH11357. 


\vspace{-2mm}


\begin{thebibliography}{5}


\ifx \showCODEN    \undefined \def \showCODEN     #1{\unskip}     \fi
\ifx \showDOI      \undefined \def \showDOI       #1{#1}\fi
\ifx \showISBNx    \undefined \def \showISBNx     #1{\unskip}     \fi
\ifx \showISBNxiii \undefined \def \showISBNxiii  #1{\unskip}     \fi
\ifx \showISSN     \undefined \def \showISSN      #1{\unskip}     \fi
\ifx \showLCCN     \undefined \def \showLCCN      #1{\unskip}     \fi
\ifx \shownote     \undefined \def \shownote      #1{#1}          \fi
\ifx \showarticletitle \undefined \def \showarticletitle #1{#1}   \fi
\ifx \showURL      \undefined \def \showURL       {\relax}        \fi
\providecommand\bibfield[2]{#2}
\providecommand\bibinfo[2]{#2}
\providecommand\natexlab[1]{#1}
\providecommand\showeprint[2][]{arXiv:#2}

\bibitem[\protect\citeauthoryear{Chakraborty, Subramoni, and Panda}{Chakraborty
  et~al\mbox{.}}{2017}]%
        {chakraborty2017contention}
\bibfield{author}{\bibinfo{person}{Sourav Chakraborty}, \bibinfo{person}{Hari
  Subramoni}, {and} \bibinfo{person}{Dhabaleswar~K Panda}.}
  \bibinfo{year}{2017}\natexlab{}.
\newblock \showarticletitle{Contention-aware kernel-assisted MPI collectives
  for multi/many-core systems}. In \bibinfo{booktitle}{\emph{2017 IEEE
  International Conference on Cluster Computing (CLUSTER)}}. IEEE,
  \bibinfo{pages}{13--24}.
\newblock


\bibitem[\protect\citeauthoryear{Hashmi, Chakraborty, Bayatpour, Subramoni, and
  Panda}{Hashmi et~al\mbox{.}}{2018}]%
        {hashmi2018designing}
\bibfield{author}{\bibinfo{person}{Jahanzeb~Maqbool Hashmi},
  \bibinfo{person}{Sourav Chakraborty}, \bibinfo{person}{Mohammadreza
  Bayatpour}, \bibinfo{person}{Hari Subramoni}, {and}
  \bibinfo{person}{Dhabaleswar~K Panda}.} \bibinfo{year}{2018}\natexlab{}.
\newblock \showarticletitle{Designing efficient shared address space reduction
  collectives for multi-/many-cores}. In \bibinfo{booktitle}{\emph{2018 IEEE
  International Parallel and Distributed Processing Symposium (IPDPS)}}. IEEE,
  \bibinfo{pages}{1020--1029}.
\newblock


\bibitem[\protect\citeauthoryear{Hori, Si, Gerofi, Takagi, Dayal, Balaji, and
  Ishikawa}{Hori et~al\mbox{.}}{2018}]%
        {hori2018process}
\bibfield{author}{\bibinfo{person}{Atsushi Hori}, \bibinfo{person}{Min Si},
  \bibinfo{person}{Balazs Gerofi}, \bibinfo{person}{Masamichi Takagi},
  \bibinfo{person}{Jai Dayal}, \bibinfo{person}{Pavan Balaji}, {and}
  \bibinfo{person}{Yutaka Ishikawa}.} \bibinfo{year}{2018}\natexlab{}.
\newblock \showarticletitle{{Process-in-Process: Techniques for Practical
  Address-Space Sharing}}. In \bibinfo{booktitle}{\emph{Proceedings of the 27th
  International Symposium on High-Performance Parallel and Distributed
  Computing}}. ACM, \bibinfo{pages}{131--143}.
\newblock


\bibitem[\protect\citeauthoryear{Huang, Di, Yu, Zhai, Liu, Raffenetti, Zhou,
  Zhao, Chen, Cappello, Guo, and Thakur}{Huang et~al\mbox{.}}{2023}]%
        {huang2023ccoll}
\bibfield{author}{\bibinfo{person}{Jiajun Huang}, \bibinfo{person}{Sheng Di},
  \bibinfo{person}{Xiaodong Yu}, \bibinfo{person}{Yujia Zhai},
  \bibinfo{person}{Jinyang Liu}, \bibinfo{person}{Ken Raffenetti},
  \bibinfo{person}{Hui Zhou}, \bibinfo{person}{Kai Zhao},
  \bibinfo{person}{Zizhong Chen}, \bibinfo{person}{Franck Cappello},
  \bibinfo{person}{Yanfei Guo}, {and} \bibinfo{person}{Rajeev Thakur}.}
  \bibinfo{year}{2023}\natexlab{}.
\newblock \bibinfo{title}{C-Coll: Introducing Error-bounded Lossy Compression
  into MPI Collectives}.
\newblock
\newblock
\showeprint[arxiv]{2304.03890}~[cs.DC]


\bibitem[\protect\citeauthoryear{Parsons and Pai}{Parsons and Pai}{2014}]%
        {parsons2014accelerating}
\bibfield{author}{\bibinfo{person}{Benjamin~S Parsons} {and}
  \bibinfo{person}{Vijay~S Pai}.} \bibinfo{year}{2014}\natexlab{}.
\newblock \showarticletitle{Accelerating MPI collective communications through
  hierarchical algorithms without sacrificing inter-node communication
  flexibility}. In \bibinfo{booktitle}{\emph{2014 IEEE 28th International
  Parallel and Distributed Processing Symposium}}. IEEE,
  \bibinfo{pages}{208--218}.
\newblock


\end{thebibliography}


\end{document}